\documentclass[amsmath,amssymb, aps, prl, superscriptaddress]{revtex4-2}
\usepackage{float}
\usepackage{graphicx}
\usepackage{dcolumn}
\usepackage{bm}
\usepackage{hyperref}
\hypersetup{
    colorlinks=true,
    linkcolor=blue,
    filecolor=magenta,      
    urlcolor=cyan,
}
\usepackage{color}
\usepackage{fancyhdr}
\usepackage[normalem]{ulem}

\newcommand{\risp}[1]{\textcolor{black}{{#1}}}
\newcommand{\rizp}[1]{\textcolor{black}{{#1}}}

\begin{document}

\title{Detecting milli-Hz gravitational waves with optical resonators}

\author{Giovanni Barontini}
\email{g.barontini@bham.ac.uk}
\affiliation{School Of Physics and Astronomy, University of Birmingham, Edgbaston, Birmingham, B15 2TT, UK}
\author{Xavier Calmet}
\affiliation{Department of Physics and Astronomy, University of Sussex, Brighton BN1 9QH, UK}
\author{Vera Guarrera}
\affiliation{School Of Physics and Astronomy, University of Birmingham, Edgbaston, Birmingham, B15 2TT, UK}
\author{Aaron Smith}
\affiliation{School Of Physics and Astronomy, University of Birmingham, Edgbaston, Birmingham, B15 2TT, UK}
\author{Alberto Vecchio}
\affiliation{School Of Physics and Astronomy, University of Birmingham, Edgbaston, Birmingham, B15 2TT, UK}

\date{\today}

\begin{abstract}
We propose a gravitational wave detector based on ultrastable optical cavities enabling the detection of gravitational wave signals in the mostly unexplored $10^{-5}-1$ Hz frequency band. We illustrate the working principle of the detector and discuss that several classes of gravitational wave sources, both of astrophysical and cosmological origin, may be within the detection range of this instrument. Our work suggests that terrestrial gravitational wave detection in the milli-Hz frequency range is potentially within reach with current technology. 
\end{abstract}

\maketitle

\section{introduction}

The development of instruments capable of detecting gravitational waves (GWs) has been a notoriously lengthy and tortuous journey, since  Einstein's theoretical prediction dating over a century ago~\cite{Einstein:1916, Einstein:1918}. Because astrophysical and cosmological sources of GWs have a rich spectrum covering many decades in frequency -- from $\sim 10^{-18}\,\mathrm{Hz}$ to $\sim 1\,\mathrm{GHz}$ -- different techniques and technologies have been used or proposed to target specific frequency regimes. 

At present, three frequency regimes are accessible, each providing information about different classes of sources, and a variety of astrophysical and cosmological processes. At high frequency $\sim 10\,\mathrm{Hz} - 1\,\mathrm{kHz}$, the network of ground-based laser interferometers {of the LIGO-Virgo-KAGRA collaboration~\cite{aLIGO,aVirgo,10.1093/ptep/ptaa125}}  
are carrying out regular observing runs at increasing sensitivity level{, corresponding currently to a GW characteristic amplitude of $\sim 10^{-22}$ at $\sim 100\,\mathrm{Hz}$.} Since the first detection of GWs produced by the merger of a binary black hole~\cite{GW150914-DETECTION}, approximately 90 events 
have been observed~\cite{O1:BBH, GW170817-DETECTION, GWTC-1, GWTC-2, GWTC-3}. Major new facilities for a next generation observatory, e.g. Einstein Telescope~\cite{Punturo:2010zz} and  Cosmic Explorer~\cite{2021arXiv210909882E}, are under study.
At ultra-low frequencies $\sim 1-10\,\mathrm{nHz}$, pulsar timing arrays~\cite{FosterBacker:1990, 1979ApJ...234.1100D} have been collecting data for over twenty years{, reaching a GW strain sensitivity around $10^{-15}$}. This effort has recently provided some tentative 
evidence for a GW stochastic background
\cite{EPTA-DR2-SGWB, NANOGrav-15yr-SGWB, PPTA-DR3-SGWB, CPTA-DR1-SGWB}.  
At extremely-low frequency $\sim 10^{-15} - 10^{-18}\,\mathrm{Hz}$,
 cosmic microwave background (CMB) polarisation experiments target indirectly the GW stochastic background through the B-mode signature on the polarisation maps~\cite{1997PhRvL..78.2058K, 1997PhRvL..78.2054S}. So far, only upper-limits have been placed, see e.g.~\cite{2016ARA&A..54..227K} and references therein.

The milli-Hz frequency range $\sim 10^{-5} - 1\,\mathrm{Hz}$ (``mid-band'', thereafter) is a rich window of astrophysical and cosmological sources, which is essentially unexplored. So far, the only GW searches have been carried out using the Doppler tracking of interplanetary space-probes~\cite{Estabrook:1975jtn, Anderson:1984yh, Anderson:1993aq, Bertotti:1998rj, Armstrong:2003ay}, {satellite laser ranging measurements \cite{diegoPhysRevLett.128.101103}, or by monitoring the normal modes of the Earth \cite{PhysRevD.90.042005}, all resulting in upper-limits}. The importance of this frequency range is testified by the many different detection technologies currently under development. These range from space-borne laser interferometers such as LISA \cite{LISA2017pwj, 2024arXiv240207571C}, Taiji~\cite{ruan2020taiji} and TianQuin~\cite{2021PTEP.2021eA107M}, to terrestrial \cite{abend2023terrestrial} and space-borne \cite{aedge} atom interferometers, and optical atomic clocks on satellites~\cite{PhysRevD.94.124043}. 
All these observatories will have sensitivity in the frequency window $\sim 10^{-4} - 1\,\mathrm{Hz}$, targeting several sources: galactic and local group stellar-mass compact binaries of (primarily) white dwarfs, neutron stars and stellar-mass black holes~\cite{2018A&A...619A..53T, 2019MNRAS.483.5518K}, including guaranteed sources already known by other means~\cite{2024ApJ...963..100K, 2022ApJ...937..118W, 2023MNRAS.522.5358F}, the coalescence of massive black hole binaries, see e.g.~\cite{2021NatRP...3..732V} and references therein, the in-spiral of stellar-mass compact objects onto a massive black hole~\cite{2004PhRvD..69h2005B}, and possibly backgrounds from the early universe~\cite{2000PhR...331..283M}. 

The novel detector concept that we introduce here leverages significant advances in the technology of optical resonators, driven by the development of optical atomic clocks  ~\cite{RevModPhys.87.637,abdelhafiz2019guidelines}. In this work, we first demonstrate the working principle of a GW detector based on monolithic optical cavities. \rizp{The essential point of our approach is that gravitational waves in the mid-band do not deform the rigid spacer of the cavity, whose eigenmodes lie at kHz frequencies, but instead alter the phase of the light as it propagates between the mirrors. The cavity length defined by the spacer therefore remains constant, while the optical path through space varies with the passing wave.} We show that state-of-the-art optical cavities have sufficient sensitivity to be readily employed to search for GWs in the mid-band. We discuss the science reach enabled by networking two or more such detectors. The discovery space includes potential detection of galactic stellar-mass compact binaries and double degenerate type IA supernovae, massive black hole mergers up to cosmological distances and the probing of the stochastic GW background down to the $10^{-6}$ limit. Future programs such as {the LISA and Taiji} space-borne missions will feature orders of magnitude better sensitivity in this frequency range, but their science operation is more than a decade away, leaving ample room for discoveries in an essentially pristine parameter space, using the detectors proposed here. 

\section{Principle of Operation} 

GWs can be detected by measuring the phase difference $\Delta\varphi(t)$ between laser beams travelling along two paths that are affected differently by the passing of a GW \cite{Thorne}. The associated phase variation is {$\Delta\varphi(t)=kh(t)\Delta \mathcal{L}/2$}, where $h(t)$ is the GW strain amplitude, $\Delta \mathcal{L}$ is the distance along the two paths that is affected differently by the GW, 
and $k$ is the magnitude of the wavevector of the light. 
Laser interferometers currently in operation detect GWs by looking at the relative change of phase $\Delta\varphi(t)$ between the light travelling along their two perpendicular arms. 
As we explain below, the detector proposed here measures $\Delta\varphi(t)$ in two different ways. One is between the light propagating along two perpendicular ultrastable cavities, the other is between the light travelling inside an ultrastable cavity and an atomic reference.

Recent progress in optical clock technology enable us to measure with high precision the phase and frequency of laser light ~\cite{RevModPhys.87.637,abdelhafiz2019guidelines}. 
Clock transitions are nowadays interrogated with lasers that have linewidths approaching the ones of the transitions themselves ($\simeq \mathrm{mHz}$). To achieve such narrow linewidths the lasers are frequency- and phase- stabilized to the resonance of a monolithic optical cavity, so that the fractional cavity length stability is transferred to the fractional frequency stability of the laser as $\Delta l/l=\Delta f/f$, which is related to phase variations by $2\pi\Delta f(t)=d[{\Delta\varphi(t)}]/dt$. With this arrangement ultrastable lasers can achieve a fractional frequency instability below $10^{-16}$ ~\cite{Hafner:15,Schioppo2022, PhysRevLett.118.263202,Kedar}.

State-of-the-art optical Fabry-Perot cavities are usually composed by two high-reflective mirrors passively kept at a fixed distance by a spacer made by ultralow expansion material \cite{Hafner:15,Kedar}. \risp{As discussed with more details later,} these cavities are normally $\leq1$ m long and can be extremely well isolated from the environment both mechanically and thermally. Ultrastable cavities can in fact operate in the $10^{-3} - 1\,\mathrm{Hz}$ band at the fundamental limit, given by the Brownian thermal noise of the mirror's high-reflectivity coating ~\cite{Kedar,PhysRevLett.118.263202,zuzzoPhysRevX.13.041002}. Due to their compact size, they are marginally sensitive to Newtonian and seismic noise, which are the main sources of noise for km-scale interferometers in this frequency band (and the reason why current programmes in the  mid-band involve space missions). At frequencies below 1 mHz, the length of ultrastable cavities varies due to relaxation and creep processes characteristic of the material used for the spacer that sets the distance between the mirrors. The drift dynamics can be however highly predictable, allowing for compensation down to frequencies $\sim 10^{-5}$ Hz. In summary, due to their exceptional fractional stability, state-of-the art utrastable cavities are excellent candidates for the detection of GWs in the $10^{-5}-1$ Hz frequency range.

\subsection*{Effect of GWs on the ultrastable cavity}

 Let us recall that GWs perturb the Riemann tensor, which describes space-time curvature, but do not exert forces on free-falling test masses, as this would violate the equivalence principle. However, the time-varying curvature affects extended objects, causing them to experience time-dependent tidal forces as they try to adapt to the changing Riemann tensor (see, e.g., \cite{Thorne,Thorne2,Thorne3, Schutz, Schutz_2022}).

An extended object connecting two free-falling test masses or -equivalently- any form of interaction between two test masses  is often modeled with a spring connecting them. In absence of gravity, the motion of the spring under the influence of GWs is described by a driven damped harmonic oscillator. The drive is the change of the space-time curvature coming from the GW, the damping coefficient and the characteristic frequency of the spring are given by the properties of the object connecting the two test masses, see e.g. \cite{Thorne,Thorne2,Thorne3, Schutz, Schutz_2022}. If we consider the GW strain $h(t)=h\sin(\omega t)$ to be aligned with the spring, where $\omega=2\pi f$, and ignoring damping effects, the equation for the stretching and compressing of the spring in its reference frame is:
\begin{equation}
    \frac{d^2\epsilon}{dt^2}+\Omega^2\epsilon=\frac{l}{2}\frac{d^2h(t)}{dt^2}=-\frac{l\omega^2}{2}h\sin(\omega t),
\end{equation}
where $\Omega$ is the characteristic frequency of the spring, and $l$ is its length. The meaning of such an equation is often misunderstood. It is crucial to remark that this is not the equation of motion of the test masses, but of the stretching and compressing of the spring connecting them. This is apparent by looking at the well known (stationary) solutions:
\begin{equation}
    \epsilon(t)=\frac{l\omega^2}{2(\omega^2-\Omega^2)}h\sin(\omega t).
    \label{eq_epsilon}
\end{equation}

To understand the effect of GWs on an optical cavity cavity, it is useful to discuss two limits:
\begin{itemize}
\item $\Omega \ll \omega$ or  $\Omega \rightarrow0$. This includes the case of vacuum or vanishing interaction between the masses.  This is the situation better describing LIGO-like detectors. The ‘weak spring’ therefore represents the light’s geodesic path in the vacuum. In this case the length of the `spring' is (sinusoidally) modified by a quantity $hl/2$. GWs change the space-time curvature between the masses, that in this case corresponds to stretching and compressing a `spring' with vanishing strength, that is unable to push or pull them (in line with what discussed before). The physical interpretation is clear: vanishing interactions modified by GWs are still vanishing interactions and result in no force on the test masses (think about a limp spring).

\item $\Omega\gg\omega$ This is the case of a rigid object between the two test masses, or strong interaction between the masses. This is the situation better describing the ultra-low expansion spacer. For such spacers the lowest eigenmodes are in the tens of kHz range. As obvious from Eq. \ref{eq_epsilon}, the effect of milli-Hz GWs on such a rigid object is negligible. Also in this case, GWs don't pull or push the test masses at the ends of the spring.
\end{itemize}

The mirrors in an optical cavity are effectively connected by both kinds of `springs': one very weak (the vacuum between the mirrors) and one very strong (the ultra-low expansion spacer). In the milli-Hz range, neither these springs exert a force on the mirrors of the cavity that is driven by GWs. When a GW arrives, it changes the space-time curvature between the mirrors but not the length of the spacer connecting them. In other words, it changes the \emph{optical distance} between the mirrors but not the \emph{length} of the cavity. Therefore, in this frequency range an optical cavity cannot be used as a bar detector to measure the (vanishing) change in the length of the spacer. However, in analogy with light-interferometer detectors, it can be used to measure the change in the path traveled by the light in the vacuum between the two mirrors, which is dictated by the varying space-time curvature.

We finally note that when a spring connects two test masses, one has in principle to consider the vibrational motion, that could be for example thermally activated. However, the first case detailed above is often called `quasi free-falling'. This is because the vibrational motion induced by the spring is much slower than the timescale of interest for the detector. The motion of the test masses is quasi-stationary, so it can be ignored. In the second case, the vibrational motion induced by the spring is instead very fast and is unaffected by the passage of GWs. This can produce a modulation of the cavity length at frequency $\Omega$, that however averages to zero on the larger timescales of interest for GW detection, $1/\omega$. In both cases, it is therefore safe to ignore the effect of the normal mode vibrations.

\subsection*{The working principle}

The working principle of our detector can be understood utilizing a gauge-independent framework based on physically measurable observables. In contrast with gauge-dependent derivations, such an approach does not leave space to interpretation. To this end, we follow the treatment of \cite{xczPhysRevD.90.062002,ttPhysRevD.88.082003}, where it is demonstrated that the phase shift in a light detector can be directly related to the Riemann tensor, which is the only meaningful physical observable, because it is gauge-invariant. 
 
In the following we ignore the normal mode vibrations for the reasons discussed above. Without loss of generality, we choose the mirrors to be aligned along the axis of polarisation of an incoming GW. In the reference frame of the centre of mass of the cavity, the effect of such a GW on the phase of the light that travels from one mirror to the other can be calculated integrating the GW contribution to the Riemann curvature along the null geodesic \cite{xczPhysRevD.90.062002}:
\begin{equation}
    \frac{d^2\varphi(t)}{dt^2}=-\frac{\omega}{c}\int_0^{\lambda_1}R^x_{0x0}d\lambda=\frac{\omega}{2c}\int_0^{\lambda_1}\frac{d^2h[t(\lambda)]}{dt^2}d\lambda,
    \label{eqphase}
\end{equation}
where $R_{\alpha\beta\gamma\delta}$ is the Riemann tensor, $\lambda$ is an affine parameter distance along the null geodesic, and $\lambda_1$ is its value at the position of the second mirror. Note that the above result is gauge-independent, and all quantities, such as the worldlines of the two mirrors and the null geodesic, are evaluated in the local frame of the centre of mass of the cavity (Minkowski background), in the absence of GWs. The identity $-R^x_{0x0}=\ddot{h}/2$ can be demonstrated using the TT gauge, see e.g. \cite{Thorne}.

For a typical cavity, the time of flight of a photon from one mirror to the other is on the order of a few ns. For GWs in the milli-Hz band, we can consider the Riemann tensor as constant during this time of flight, i.e., $h[t(\lambda)]=h(t)$, therefore Eq. (\ref{eqphase}) simplifies as:
\begin{equation}
    \frac{d^2\varphi(t)}{dt^2}=\frac{\omega l}{2c}\frac{d^2{h}(t)}{dt^2},
\end{equation}
from which it follows that
\begin{equation}
    \frac{d\varphi(t)}{dt}=\frac{\omega l}{2c}\frac{dh(t)}{dt},
    \label{eq_main}
\end{equation}
This is a central result of our work, that shows that the phase of the light traveling inside the cavity is a measure of the strain induced by incoming GWs. 

If we now enable the photon to be reflected by the second mirror and come back along the same path, we can compute the variation of the phase after a round trip:
\begin{equation}
    \frac{d\varphi(t)_{RT}}{dt}=\frac{\omega l}{c}\frac{dh(t)}{dt}.
\end{equation}
By using a high finesse optical cavity we can reflect the light back and forth $n\gg1$ times without appreciable losses, so that 
\begin{equation}
    \frac{d\varphi(t)_{Cavity}}{dt}=\frac{\omega L}{2c}\frac{dh(t)}{dt}.
    \label{eq_cavity}
\end{equation}
This corresponds to `folding' $n$ times inside the cavity the total distance travelled by the light $L=2nl$ (as done for example in LIGO), and at the same time keeping the detector compact enough so that it can be effectively isolated from external noise sources. 
Typically, the cavity enhancement allows the light to travel for $L\simeq$ tens of km within a detector of only $l\leq1$~m \cite{Kedar,Hafner:15}.

In practice, to actually measure the quantity in Eq. \ref{eq_cavity} one needs to have a reference. The best choice for this reference is another laser beam at the same frequency, that has travelled a different path. Indeed if we `beat' the light that has travelled in the cavity and the reference by sending them together to a photodiode, we can measure:
\begin{equation}
   \frac{d \Delta\varphi(t)}{dt}=\frac{\omega\Delta\mathcal{L}}{2c}\frac{dh(t)}{dt},
\end{equation}
with $\Delta\mathcal{L}$ the distance along the two paths that has been affected differently by the GW. The corresponding phase variation between the light in the cavity and the reference is therefore
\begin{equation}
    \Delta \varphi(t)=k\Delta\mathcal{L}h(t)/2.
\label{eqs7}
\end{equation}
The GW strain can thus be measured by measuring $\Delta\varphi/\varphi=h(t)\Delta\mathcal{L}/2L$ between the light that exits the cavity and a reference. The precise measurement of $\Delta\varphi/\varphi$ is the aim of optical frequency metrology. 

Note that the exact same results can be obtained using gauge-dependent approaches, for example the use of the TT gauge is rather straightforward \cite{Schutz_2022, Schutz}. 

\section{The milli-Hz GW detector}

\begin{figure*}
\centering
\includegraphics[width=0.6\textwidth]{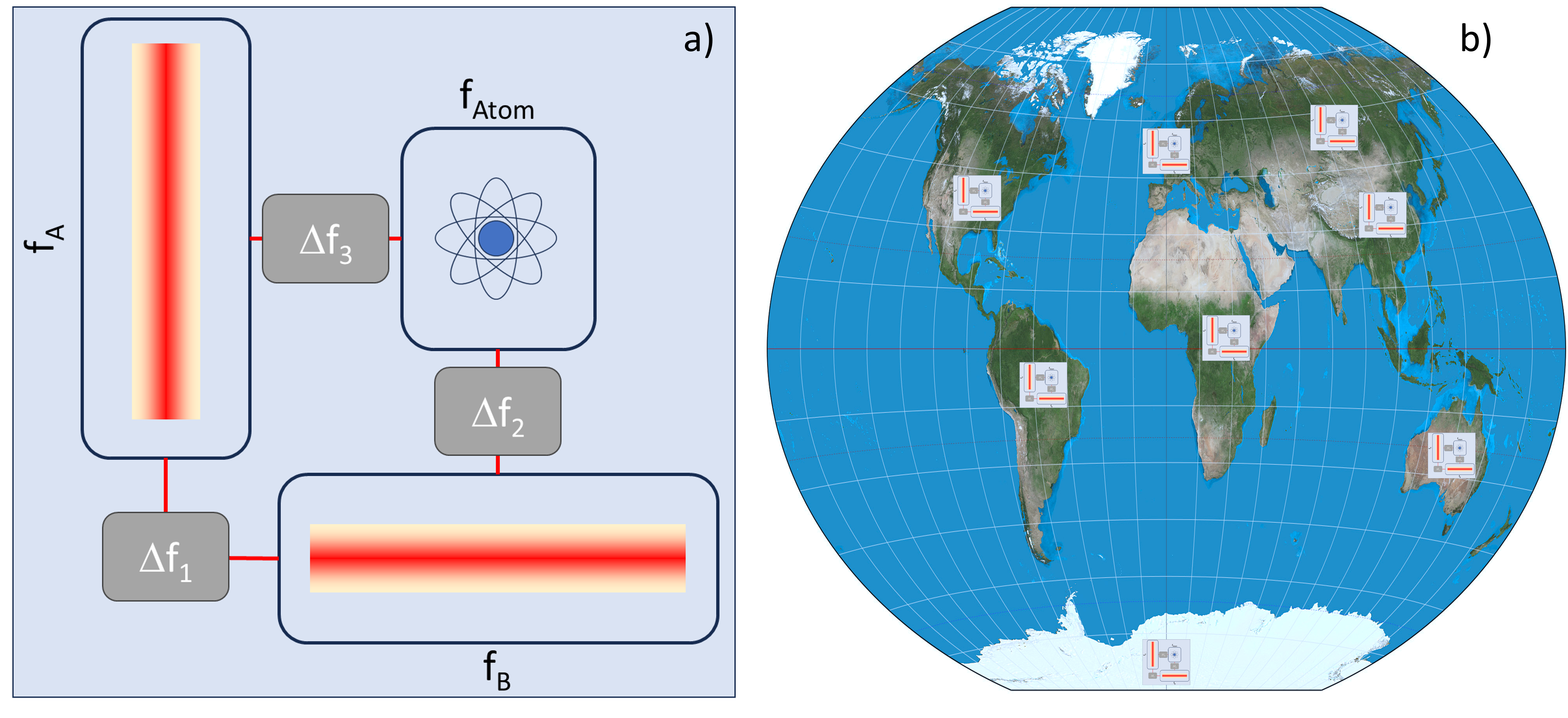}
\caption{a) Schematic of the proposed GW detector. The detector consists of two ultrastable lasers {at frequencies $f_\text{A}$ and $f_\text{B}$} and one optical atomic reference {at frequency $f_{\text{atom}}$}. The detection channels $\Delta f_i$ can be realized by beating the laser beams. b) A global network of detectors in the milli-Hz range} 
\label{Fig1}
\end{figure*}

From Eq. \ref{eqs7} it follows that a good choice for the reference could be a beam of light that has travelled for a short distance $d\ll L$, so that $\Delta\mathcal{L}\simeq L$. A better choice is a laser beam that has travelled along a perpendicular identical cavity, so that if $h(t)\equiv h_+(t)$ (polarization aligned along the axes of the two cavities), $\Delta\mathcal{L}= 2L$. {Following these considerations, we propose a detector with the configuration shown in Fig. \ref{Fig1}a). This comprises two lasers locked on orthogonal ultrastable cavities {(providing the short-term stable frequencies $f_{\mathrm{A}}$ and $f_{\mathrm{B}}$, respectively)}, and an optical atomic frequency reference {(with the long-term stable frequency $f_{\mathrm{atom}}$)}. As shown in Fig. \ref{Fig1}, $\Delta f(t)$ and therefore $\Delta\varphi(t)$ can be measured using three measurement channels: (1) comparison of two orthogonal ultrastable cavities ($\Delta f_1\equiv f_{\mathrm{A}}-f_{\mathrm{B}}$), (2) the cavity A against the atomic reference ($\Delta f_2 \equiv f_{\mathrm{A}}-f_{\mathrm{atom}}$), (3) the cavity B against the atomic reference ($\Delta f_3 \equiv f_{\mathrm{B}}-f_{\mathrm{atom}}$).} 

Multiple measurement channels in one system enable GWs to be detected with greater confidence. A positive detection of a GW signal is confirmed by the observation of correlated signals in at least two measurement channels, depending on the GW polarisation. {As an example, in the case $h(t)\equiv h_+(t)$ our detector will measure $\Delta\varphi_1/\varphi=h(t)$ and $\Delta\varphi_2/\varphi=-\Delta\varphi_3/\varphi=h(t)/2$. For the case $h(t)\equiv h_\times(t)$ instead (polarization at 45 degrees with respect to the axes of the two cavities), the detector will yield $\Delta\varphi_1/\varphi=0$ and $\Delta\varphi_2/\varphi=\Delta\varphi_3/\varphi=h(t)/2\sqrt{2}$. Therefore, the configuration shown in Fig. 1 enables the detection of any GW polarisation through the correlation of the signals between the three detection channels, although with different sensitivities. Another, more complex, possible configuration} could comprise three or more cavities placed at 45 degree angles, so as to be {maximally sensitive to optimally oriented sources}, and increase detection sensitivity and confidence. One or more copies of the proposed detector would be needed at different geographical locations to fully validate any detection. Given that in the frequency band here investigated the wavelength of GWs is much larger than the size of the Earth, disseminating $N$ detectors across the globe as show in Fig. \ref{Fig1}b) would improve the detection sensitivity {to deterministic signals} by $\sqrt{N}$. 

 \subsection*{\risp{Noise sources}}
\risp{
The sensitivity of our detector to GWs in the milli-Hz band is dictated by different noise contributions. The most prominent is the \emph{Brownian thermal noise} coming from mechanical dissipation in mirror coatings, the substrates, and the spacer. This noise source causes a flicker noise floor in the cavity length and is generally the dominating term on the high frequency end of our band of interest. The overall contribution can be written as \cite{Kessler:12}:
\begin{equation}
    S_y(f)=\frac{4k_BT}{\pi f l^2} (F^{sub}+F^{coat})+S_y^{spacer}(f),
\end{equation}
where $k_B$ is the Boltzmann's constant, $T$ is the temperature, and $F$ are functions of the parameters of the materials used \cite{Kessler:12}. The spacer contribution is usually computed with finite-element analysis due to complex support geometries. From the above expression it is clear that there are two possible strategies to reduce this contribution. The first is to lower the temperature: state-of-the-art cavities use crystalline silicon spacers and substrates at cryogenic temperatures \cite{zzRobinson:19,Kedar,Kedarth}, reaching stabilities well below 1 $\times$ 10$^{-16}$. The second strategy, that also enables to reach stabilities below $10^{-16}$, consists of increasing the length of the cavity up to half a meter \cite{ffPhysRevLett.119.243601, Schioppo2022, schioppoprivate}.} 

Another important noise source in this frequency band is \emph{seismic noise}, i.e., vibrations of the cavity due to ground motion, that could cause fluctuations in the length of the cavity by mechanically coupling into the spacer and mirror positions. State-of-the-art cavities utilize a complex combination of active and passive compensation strategies and, because of their reduced size, they can reach sensitivities to vibrations below $10^{-11}/g$ \cite{zzRobinson:19, ccPhysRevA.87.023829}, where $g$ is the acceleration due to gravity. As a result, the impact of this noise source is orders of magnitude lower than the one coming from thermal noise. 

\emph{Newtonian noise} coming from gravity gradients caused by moving masses, like ground or air, could also perturb the position of the mirrors, see e.g. \cite{wwPhysRevD.30.732}. \rizp{In the long-wavelength approximation, which is relevant here, the tidal force on two independent test masses is suppressed by a factor of order $l/\lambda_{NN}$, where $\lambda_{NN}\gg l$ is the wavelength of the disturbance. Importantly, with a rigid cavity spacer, the differential length response is further suppressed by $(\Omega/\omega)^2$ (as shown earlier).} As a matter of fact, this noise source has not been observed as a limiting factor in state-of-the-art optical cavities. However, it may become relevant in the $10^{-5}$–$10^{-3}$ Hz range, if instabilities are pushed below $10^{-17}$. \rizp{As discussed earlier, the modification of the Riemann tensor induced by GWs, instead, cannot be suppressed by mechanical rigidity between the mirrors.}

\risp{Finally, \emph{slow drifts} in cavity length, due to residual thermal expansion, aging, or stress relaxation in materials, can cause a gradual frequency drift over long timescales. These drifts are highly predictable, and therefore can be compensated down to frequencies $\sim 10^{-5}$ Hz. This can be done by monitoring the individual cavities' lengths with respect to a suitable reference, such as another ultrastable cavity or an optical atomic transition \cite{Hafner:15,xxPhysRevLett.123.173201,zzRobinson:19,Kedar}. The latter, being drift-free, provides the best possible reference to make the behavior of optical cavities more predictable for frequencies  below $10^{-3}$ Hz \cite{Kedar, Kedarth}.}

\risp{In summary, the power spectral density of state-of-the-art cavities in the milli-Hz range is dominated by Brownian noise at high frequencies, and slow drifts \rizp{at} low frequencies. In this window, \emph{measured} spectra of state-of-the-art cryogenic cavities are indeed very well fitted by \cite{Kedar,Kedarth}:
 \begin{equation}
{S_y(f)}={4\times10^{-33}\text{Hz}^{-1}+1.7\times 10^{-33}f^{-1} + 9\times10^{-38}f^{-2}\text{Hz}},
\label{psdcryo}
\end{equation}
while those featuring ultra-low expansion glass spacers, are well fitted by \cite{schioppoprivate}: 
\begin{equation}
{S_y(f)}={1.1\times 10^{-34}f^{-1} + 2.2\times10^{-36}f^{-2}\text{Hz}}.
\label{psdule}
\end{equation}
In both cases, the fit corresponds to the calculated contributions coming from the thermal noise and the slow drift.}

\section{The science case}

\begin{figure*}
 \centering
 \includegraphics[width=0.8\textwidth]{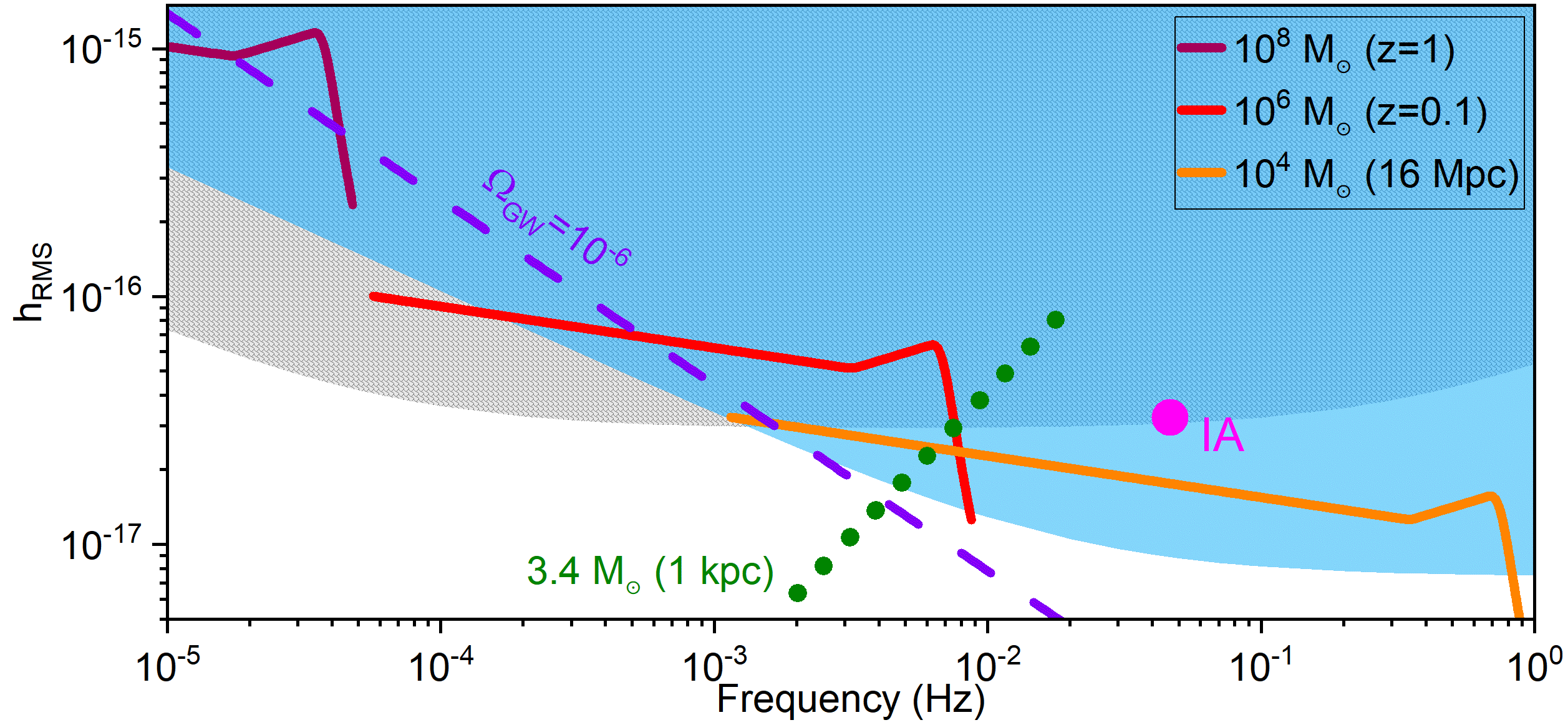}
 \caption{{Strain sensitivity $h_{\text{RMS}}=\sqrt{S_y(f)f}$ as a function of the frequency for a detector comprising two orthogonal cavities and an atomic reference, compared to GW signals from several source classes. The light blue shaded area is the parameter space accessible with ultra-low expansion glass cavities, while the patterned area with cryogenic silicon cavities. The continuous lines are the characteristic strains produced by the coalescence of optimally orientated binaries of equal mass, non-spinning black holes. The green dotted line corresponds to the quasi-monochromatic characteristic amplitude produced by a galactic stellar mass binary black hole with chirp mass ${\cal M} = 3.4$ M$_{\odot}$ at a distance of $D=1\,\mathrm{kpc}$ and observed for $T_\mathrm{obs}=3$ years. The magenta dot shows the characteristic amplitude of a double white-dwarf progenitor of a type IA supernova at 1 kpc. The dashed line is the effective characteristic strain produced with signal-to-noise ratio of 1 by a stochastic background with an energy density $\Omega_{GW}= 10^{-6}$ assuming the correlation of the data streams of ten optimally orientated detectors for $T_\mathrm{obs}=3$ years. See \cite{SuppMat} for details.}}
 \label{Fig2}
 \end{figure*}

\risp{We summarize in Fig. \ref{Fig2} the science reach of the proposed detector. The patterned area is the sensitivity to GW signals of a detector based on an orthogonal pair of cryogenic silicon cavities whose power spectral density is given by Eq. (\ref{psdcryo}), while the blue shaded area is for a detector based on cavities featuring ultra-low expansion glass spacers, whose power spectral density is given by Eq. (\ref{psdule}).}

For galactic sources, we show the characteristic amplitude of a quasi-monochromatic signal from a stellar-mass binary black hole (green dotted line) with chirp mass of ${\cal M} = 3.4$  M$_\odot$ (corresponding to the median value of the expected population according to recent models~\cite{2022ApJ...937..118W}) at 1 kpc distance. 
We also show the GW signal associated with a double degenerate type IA supernova (magenta dot) at 1 kpc \cite{Mandel_2018}. The observation of a GW signal which is coincident in time and sky position with a supernova would be a strong indication that the source of the supernova is the merger of two white dwarfs \cite{10.1111/j.1365-2966.2004.07713.x}. The solid curves in Fig. \ref{Fig2} show the last four months of the signals that would be produced by 
the coalescence of extra-galactic binary black holes with different masses and distances. Our proposed detector would be able to observe the coalescence of a binary black hole with ${\cal M} = 10^4$  M$_\odot$ at the distance of the Virgo cluster. Many astrophysical models predict the existence of these intermediate mass binary black holes, see, e.g., \cite{annurev1,millerrev,Merritt_2005} and references therein. They are however extremely challenging to search for, and consequently there is no evidence to date for these sources, and their merger rates are highly uncertain. Our proposed detector would also be able to detect the merger of massive black holes at cosmological distances. These events are predicted to be rare: $\approx 10^{-6}-1$ yr$^{-1}$, using merger rates consistent with the recent pulsar timing array observations.
Finally, the purple dashed line corresponds to the signal that a stochastic GW background with energy density $\Omega_{GW}=10^{-6}$ would produce on a network of 10 detectors. 
In general \footnote{This is not the case for the most massive binaries within the detector sensitivity range} the GW sources in this frequency band are long-lived compared to the timescale of 24 hours ({with some of them living longer than} one year) over which the detector changes location and orientation. The detector motion therefore introduces source-location dependent modulations in amplitude and phase to the signal detected. This characteristic modulation enables the determination of the source location in the sky, and provides additional power to discriminate between a GW signal and instrumental noise. 

\section{conclusions}

In summary, this work opens the window for terrestrial gravitational wave detection in the mid-band, and makes the case for a global network of optical frequency metrology detectors as a precursor for the large-scale observatories and space-borne missions that will become operative in the next decades. We have proposed a detector utilizing a combination of two or more ultrastable cavities and an optical atomic reference. These tabletop detectors are relatively compact and easy to implement and maintain, making them ideal for global dissemination. This could also be facilitated by the infrastructure developed or under development for networks of clocks \cite{collaboration2021frequency,Schioppo2022, ecfa, barontini2022measuring}. We have discussed the science reach of the detector based on two different types of optical resonators. Compact binaries in the Milky Way could be detected in the milli-Hz range, as well as black hole mergers at cosmological distances. Both positive and null detection will shine some light on the population of these elusive sources. The magnitude of the gravitational wave background could also be probed beyond cosmological indirect limits for frequencies below 10$^{-2}$ Hz. 

Together with global dissemination, further improvements in the technology for optical frequency metrology, pushed also by the next redefinition of the second,  will enable the probing of increasingly larger regions of the GW parameter space. The extension of the methodologies proposed here to dark fibre networks connecting atomic clocks could enable terrestrial GW detection at lower frequencies. The possibility of a joint detection between the network of laser interferometers and a network of mid-band detectors is especially intriguing to gain new insight into binary systems over an unprecedented frequency range.   

\paragraph*{Acknowledgements}
We acknowledge fruitful discussions with the members of the QSNET consortium. We thank A. Giusti, R. Godun, T. Easton, M. Favier, I. Hill, S. Hsu, M. Keller, D. Martynov, A. Parke, G. Prattern, L. Prokhorov, M. Schioppo, and M. Tarbutt for useful discussions.
This work was supported by STFC and EPSRC under grants ST/T006048/1,
ST/T00603X/1, ST/T00102X/1, ST/W006138/1, ST/Y00454X/1, ST/Y004418/1. AV acknowledges the support of the Royal Society and Wolfson Foundation.

\newpage
\section{Appendix}

\subsection*{Sources of gravitational waves in the mid-band}

\subsubsection*{Binary systems}

We consider binary systems of compact objects of individual source-frame masses $m_{1,2}^{(\mathrm{src})}$ in circular orbit at (luminosity) distance $D$. We indicate the observed (redshifted, or detector-frame) mass parameter $m$ as $m = m^{(\mathrm{src})} (1 +z)$, where $z$ is the redshift. Due to emission of gravitational waves a binary loses energy and, as consequence, the orbital separation, $a$ decreases. At the leading Newtonian order the orbital separation is related to the gravitational wave frequency, $f$, as
\begin{eqnarray}
    f & = & 0.2\,\left[\frac{M^{(\mathrm{src})}\,(1 + z)}{10^7\,M_\odot}\right]^{1/2}\,
    \left[\frac{a}{1\,\mathrm{AU}}\right]^{-3/2}\,\mathrm{mHz}\,
    \nonumber \\
    & = & 0.2\,\left[\frac{M}{1\,M_\odot}\right]^{1/2}\,
    \left[\frac{a}{1\,R_\odot}\right]^{-3/2}\,\mathrm{mHz}
    \label{eqs1}
\end{eqnarray}
In the previous expression $f$ is the \textit{observed} (detector-frame) frequency, and $M = m_1 + m_2$ the detector frame total mass. 
Once the binary reaches the innermost stable circular orbit, whose associated GW frequency is 
\begin{equation}
    f_\mathrm{ISCO}^{(2,2)} \sim 0.4\,\left[\frac{M^{(\mathrm{src})}\,(1 + z)}{10^7\,M_\odot}\right]^{-1}\,\mathrm{mHz}\,.
\end{equation}
the binary starts to merge and settles into a remnant black hole with dominant quasi-normal mode frequency 
\begin{equation}
    f_\mathrm{ring}^{(2,2)} \sim 3.2\,\left[\frac{M^{(\mathrm{src})}\,(1 + z)}{10^7\,M_\odot}\right]^{-1}\,\mathrm{mHz}\,.
\end{equation}

The optimal signal-to-noise ratio (SNR) for a certain signal is given by \cite{PhysRevX.6.041015}:
\begin{equation}
    \text{SNR}=\sqrt{\int_0^\infty 4\frac{|\tilde{h}(f)|^2}{S_y(f)}df},
    \label{eqsnr}
\end{equation}
where $\tilde{h}(f)$ is the Fourier transform of the signal. We assume the source to be optimally located and orientated, and approximate the full coalescence signal (in-spiral, merger and ring-down) for non-spinning black holes according to the phenomenological model of Ref.~\cite{Ajith_2007,Ajith_2008,PhysRevD.77.104017}. Within this approximation, the GW signal can be written as
\begin{equation}
    |\tilde{h}(f)| = A\, G(f; m_1, m_2)
    \label{hf_PhenomA}
\end{equation}
where $G(f; m_1, m_2)$ is a function of frequency and the binary masses that during the inspiral phase reduces to the familiar $(f/f_0)^{-7/6}$, where $f_0$ is an arbitrary reference frequency. Within this approximation, the frequency domain amplitude is
\begin{eqnarray}
    A & = & \left(\frac{5}{24\pi^{2/3}}\right)^{1/2}\,
    \frac{{\cal M}^{5/6}}{D}\,f_0^{-7/6}\,\mathrm{Hz}^{-1} \nonumber
    \\
    & = & 2.5\times 10^{-12}\,\left[\frac{M^{(\mathrm{src})}\,(1 + z)}{10^7\,M_\odot}\right]^{5/6}\,
    \left[\frac{D}{1\,\mathrm{Gpc}}\right]^{-1}
    \left[\frac{f_0}{0.1\,\mathrm{mHz}}\right]^{-7/6}\,\mathrm{Hz}^{-1}\,,
    \label{eqnA}
\end{eqnarray}
where ${\cal M} = \mu^{3/5}M^{2/5}$ is the chirp mass and $\mu = m_1 m_2/M$ the reduced mass.

At the leading post-Newtonian order the time to coalescence, $\tau$ for a binary radiating at frequency $f$ is:
\begin{equation}
    \tau = 2.6\times 10^5\, \left(\frac{f}{0.1\,\mathrm{mHz}}\right)^{-8/3}\,\left(\frac{\mu/M}{0.25}\right)^{-1}\, \left(\frac{M}{10^7\,M_\odot}\right)^{-8/3}\,\mathrm{sec}\,,
    \label{eqntau}
\end{equation}
where $\mu/M = 0.25$ corresponds to the case of an equal mass binary, $m_1 = m_2$. For massive black hole binaries, the signal is therefore a short-lived transient, compared to the time over which  the instrument operates, for days to months. 

\paragraph{Black hole mergers}
To evaluate $\tilde{h}(f)$ for a binary black hole merger, we have used the PhenomA waveform approximation \cite{Ajith_2007,Ajith_2008,PhysRevD.77.104017}, that describes the full coalescence (inspiral-merger-ringdown). For our detector, the integral in Eq. (\ref{eqsnr}) is calculated between
\begin{equation}
  f_{min}=\text{max}\left\{ \frac{1}{8\pi}\left[T_{obs}\left(\frac{G{\cal M}}{c^3}\right)^{5/3}\frac{\nu}{5}\right]^{-3/8},10^{-5} \text{Hz}\right\},
\end{equation}
where $G$ is the gravitational constant, and $\nu=0.25$ for equal mass systems, with the observation time $T_{obs}$ coinciding with the coalescence time, and 
\begin{equation}
    f_{max}=\text{min}\left\{ f_3,1 \text{Hz}\right\},
\end{equation}
where $f_3$ is the cut frequency of the waveform approximation. In Fig. 2 in the main text we show $2|\tilde{h}(f)|f$ for black hole mergers with chirp mass ${\cal M}=10^8$ M$_\odot$ at redshift $z=1$, for which we obtain a signal-to-noise ratio (SNR) of 5.4 with current cavities, and 5.5 with next generation cavities. For ${\cal M}=10^6$ M$_\odot$ at $z=0.1$, the SNR are 2.3 and 4.5 respectively. For ${\cal M}=10^4$ M$_\odot$ at 16 Mpc distance (corresponding to the distance of the Virgo Cluster), the SNR are 1.19 and 4.46, assuming the observation of the last 4 months of lifetime of the system.

\paragraph{Compact binaries}
Compact binary systems emit gravitational waves at frequency $f_0$ on the characteristic timescale, see e.g., \cite{Colpi_2017}:
\begin{equation}
    \tau=\frac{5}{\nu}(8\pi f_0)^{-8/3}\left(\frac{G{\cal M}}{c^3}\right)^{-5/3}.
\end{equation}
 For example a system with ${\cal M}=1$M$_\odot$ emits almost monochromatic radiation at 0.1 Hz for approximately 40 years. This enables long observation times $T_{obs}$ that can be used to reduce the RMS noise as
\begin{equation}
    h_{\text{RMS}}=\sqrt{S_y(f_0)\frac{1}{T_{obs}}}
    \label{eqnhrms}
\end{equation}
as long as $T_{obs}\leq\tau$. From Eq. (\ref{eqnA}), the amplitude of gravitational waves emitted by a compact binary is given by
\begin{equation}
    A(f)=2c\left(\frac{G{\cal M}}{c^3}\right)^{5/3}\frac{(f\pi)^{2/3}}{D}.
    \label{eqd}
\end{equation}
The green dotted line in Fig. 2 of the main text corresponds to the quasi-monochromatic characteristic amplitude produced by a galactic stellar mass binary black hole with chirp mass ${\cal M} = 3.4$ M$_{\odot}$ at $D=1\,\mathrm{kpc}$ and observed for $T_\mathrm{obs}=3$ years. The high-frequency end-point of the line corresponds to a source 100 years from coalescence. 

\subsubsection*{Type IA supernovae}
Together with compact binaries, this range of frequencies is of interest for the possible multi-messenger detection of a type IA supernova in our galaxy \cite{Mandel_2018}. The yellow dot in Fig. 2 in the main text shows the characteristic amplitude of a double white-dwarf progenitor of a type IA supernova at 1 kpc. This would originate from the coalescence of a white dwarf binary system with ${\cal M}=1$ M$_\odot$. The values shown in Fig. 2 in the main text are evaluated using eq. \ref{eqs1} and \ref{eqnA} considering that the minimum orbital radius for this binary system is $\simeq 0.02R_\odot$ \cite{Dan_2011}. 

\subsubsection*{Stochastic background}
A stochastic background of gravitational radiation manifests as a random gravitational wave signal stemming from a large number of weak, independent, and unresolved sources.  The energy density spectrum is described by the dimensionless quantity
\begin{equation}
    \Omega_{GW}(f)=\frac{8\pi G f}{3H_0^2c^2}\frac{d\rho_{GW}}{df}
\end{equation}
where $\rho_{GW}$ is the energy density in the frequency interval $f$ and $H_0$ is the Hubble constant. The associated power spectral density is given by \cite{MAGGIORE2000283} 
\begin{equation}
    S_h(f)=\frac{3H_0^2}{2\pi^2}\frac{\Omega_{GW}(f)}{f^3}.
\end{equation}
The SNR in Eq. (\ref{eqsnr}) for a network of $i=\{1,..,N\}$ detectors with noise power spectral densities $\{S_{y1},..,S_{yN}\}$ and a common observation time $T_{obs}$ is given by \cite{PhysRevD.59.102001,PhysRevLett.116.131102,PhysRevLett.118.151105}
\begin{equation}
    \text{SNR}={\sqrt{2T_{obs}}}\left[\int_0^\infty df\sum_{i=1}^N \sum_{j>i}\left(\frac{S_h^2(f)}{S_{yi}(f)S_{yj}(f)}\right)    \right]^{1/2}.
\end{equation}
Imposing $S_{yi}=S_{yj}=S_y$, for the case of correlating the data of $N>1$ detectors, we find a simplified expression for the SNR in the frequency band $\Delta f$:
\begin{equation}
    \text{SNR}={\sqrt{2T_{obs} \Delta f}}\sqrt{N(N-1)}\left\langle\left(\frac{S_h(f)}{S_{y}(f)}\right)^2 \right\rangle_{\Delta f}^{1/2}.
    \label{eqsnr1}
\end{equation}
The dotted line in Fig. 2 in the main text shows the effective characteristic strain produced with SNR=1 by a stochastic background with an energy density $\Omega_{GW}= 10^{-6}$ assuming the correlation of the data streams of ten optimally orientated detectors over a bandwidth $\Delta f \sim f$ for $T_\mathrm{obs}=3$ years

\bibliography{main_bibl}

\end{document}